\documentclass[12pt]{article}
\usepackage[utf8]{inputenc}

\title{OmniRemesh: Adaptive and Quasi-differentiable Remeshing for Crystal Plasticity Simulation and Inverse Parameter Calibration under Large Deformation}
\usepackage{tikz}
\usetikzlibrary{arrows.meta,positioning,shapes.geometric,fit,backgrounds}
\usepackage[numbers,sort&compress]{natbib}
\usepackage{amsmath}
\usepackage{epsfig,amsthm}
\usepackage{amssymb,amsfonts}
\usepackage{dsfont}
\usepackage{subfigure}
\usepackage{indentfirst}
\usepackage{mathrsfs}
\usepackage{color}
\usepackage{graphicx}
\usepackage{zymacros}
\usepackage{hyperref}
\usepackage{algorithm}
\usepackage{algpseudocode}
\usepackage{authblk}
\begin{document}
\author[1]{Ningyu Yan}
\author[1]{Yuntong Huang}
\author[1,2,*]{Yang Xiang}
\affil[1]{Department of Mathematics, The Hong Kong University of Science and Technology, Clear Water
Bay, Kowloon, Hong Kong, China}
\affil[2]{HKUST Shenzhen--Hong Kong Collaborative Innovation Research Institute, Shenzhen, China}

	\maketitle

\noindent\textsuperscript{*}To whom correspondence should be addressed. E-mail: maxiang@ust.hk (Y.X.)

\begin{center}
    \textbf{\large Abstract} \\[0.5em] 
\end{center}
{
\noindent Large-deformation crystal plasticity finite element method (CPFEM) simulations are often limited by accumulated mesh distortion, which degrades accuracy and numerical stability, while adaptive remeshing introduces discrete topology changes that impede gradient-based inverse analysis. We present OmniRemesh, a unified framework that addresses these forward and inverse challenges through two developments. First, a structure-driven remeshing method dynamically redistributes local mesh resolution according to both microstructural geometry and the evolving mechanical state. By refining grain boundaries and localized deformation regions while retaining a coarser mesh elsewhere, the method maintains mesh quality and physical consistency, improves the accuracy and robustness of large-deformation calculations, and resolves grain-scale heterogeneity without uniformly dense discretization. Second, a frozen-remeshing-branch strategy locally fixes the mesh sequence within a parameter trust region and periodically updates it as the parameters evolve. This treatment provides approximate automatic-differentiation sensitivities despite topology changes, enabling efficient inverse calibration of constitutive parameters against both macroscopic and local observables. Numerical examples demonstrate accurate and stable CPFEM simulations up to 80\% tensile deformation. The inverse calibration successfully recovers both macroscopic and local responses. OmniRemesh thus provides a practical framework for large-deformation CPFEM and remeshing-aware constitutive calibration.
\par}

\section{Introduction}
The crystal plasticity finite element method (CPFEM) has become an important full-field framework for investigating the mechanical response of polycrystalline materials \citep{roters2011crystal, keshavarz2013multi, sarrazola2020full, long2023meso, roters2010overview}. By resolving crystallographic orientation, anisotropic elasticity, slip-system activity, and history-dependent hardening at the grain scale, CPFEM provides a mechanistic link between microstructure and macroscopic behavior. In contrast to homogenized constitutive models, CPFEM can explicitly represent heterogeneous fields within and between grains, including stress concentrations, intragranular lattice-rotation gradients, and localized hardening near grain boundaries. It has therefore been widely applied to study stress--strain response \citep{hansen2020investigation,herrera2020influence}, texture evolution \citep{choi2010simulation,prakash2009modeling,kalidindi2009modeling}, grain interaction \citep{kanjarla2010assessment,ma2006consideration,ma2006studying}, and strain localization \citep{chakrabarty2021investigation,zan2024hydride,li2023deterministic} in metallic materials.

Realizing this full-field capability requires adequate spatial resolution. Coarse discretizations reproduce volume-averaged stress--strain behavior while missing orientation spread, grain-boundary stress concentrations, and localized slip or hardening bands. Grain-scale predictions are consequently sensitive to the number, type, shape, and arrangement of elements within individual grains \citep{zhao2007influence,lim2019investigating,weiss2024effects,feather2021numerical}. The resolution requirement becomes more difficult under medium and large plastic deformation: in a Lagrangian CPFEM formulation, heterogeneous plastic flow progressively distorts the mesh, degrades nonlinear-solver conditioning and accuracy, and can terminate the simulation prematurely \citep{quey2011large,sedighiani2021large}.

Remeshing is a practical way to maintain mesh quality, but its use in CPFEM must reconcile geometry reconstruction with the transfer of orientation- and history-dependent state variables. Existing studies have established polycrystal remeshing procedures, level-set or structured-mesh descriptions, and grain-consistent or orientation-aware state-transfer schemes \citep{resk2009adaptive,quey2011large,kim2015three,frydrych2019solution,sedighiani2021large}. Nevertheless, most approaches do not use both persistent microstructural features and evolving mechanical localization to control spatial resolution on the current deformed configuration. A large-deformation CPFEM remeshing method should therefore concentrate resolution around grain boundaries and internal-variable hot spots, preserve discontinuities between grains, and recover a mechanically admissible state after transfer.

Reliable CPFEM prediction also depends on calibrating its constitutive parameters. Quantities such as the initial slip resistance, hardening and saturation coefficients, rate sensitivity, and latent-hardening interactions are material dependent and are rarely all accessible through direct single-crystal or micromechanical tests. Inverse calibration can estimate them by minimizing discrepancies between CPFEM predictions and experimental observations. Early and widely used formulations fit one or more macroscopic stress--strain curves through local nonlinear optimization or hierarchical RVE models \citep{herreraSolaz2014inverse}. Because each objective evaluation requires a nonlinear full-field solver and the parameter landscape can be nonconvex, response-surface and genetic-algorithm approaches \citep{sedighiani2020parameters}, Gaussian-process and Bayesian optimization \citep{kuhn2022bayesian}, and surrogate-assisted sensitivity analysis \citep{dorward2024calibration} have been developed to reduce the number of expensive CPFEM evaluations. Bayesian calibration has additionally been used to quantify uncertainty and non-uniqueness in physics-based crystal-plasticity parameters \citep{nguyen2021bayesian}.

Calibration based only on macroscopic response is often insuﬀicient because CPFEM
is also expected to predict local grain-scale deformation fields. Correlated parameter combinations may give nearly indistinguishable aggregate stress--strain curves while producing different slip activity, hardening, strain localization, or lattice rotation at the grain scale. Full-field and in-situ measurements provide additional, complementary constraints: finite-element model updating informed by digital image correlation (DIC) has combined microscopic displacement fields with macroscopic loads \citep{guery2016identification}, integrated 3D-DIC has enabled joint identification of boundary conditions and crystal parameters \citep{bertin2016identification}, and recent multi-objective workflows have incorporated in-situ EBSD grain-reorientation trajectories alongside macroscopic response \citep{choi2026multiobjective}. Such local observables can improve physical identifiability, but they also increase the need for accurate spatial resolution in precisely the grain-boundary and localization regions most affected by mesh distortion.

Recent differentiable CPFEM formulations provide a complementary route to computational efficiency. Automatic differentiation supplies sensitivities of global and local objectives with respect to constitutive parameters, enabling gradient-based multi-objective calibration without finite-difference sweeps or a separately trained surrogate \citep{hu2025efficient,hu2026efficient}. Existing differentiable workflows, however, assume a fixed mesh and thus a fixed computational graph. Large-deformation remeshing breaks this assumption: changing the parameters changes the mechanical state, which can change refinement indicators, mesh connectivity, and state-transfer neighborhoods. These discrete decisions make the complete remeshing map nonsmooth and prevent direct end-to-end differentiation. The challenges of remeshing and calibration are therefore coupled rather than independent: local data demand high-resolution, deformation-following meshes, whereas the required topology updates disrupt the sensitivities needed for efficient inverse calibration.

To address these coupled challenges, we introduce OmniRemesh, whose principal contributions are twofold. First, we develop a \emph{structure-driven remeshing} method for large-deformation CPFEM. Here, ``structure'' includes both persistent microstructural geometry, represented by grain-boundary proximity, and evolving mechanical structure, represented by internal-variable indicators of slip activity and hardening. These indicators define a spatially varying target resolution on the current configuration; grain-preserving nearest-neighbor transfer and equilibrium projection then retain discontinuities and restore mechanical consistency on the new mesh. The method thereby combines mesh-quality recovery on the deformed configuration with selective resolution of grain-scale features rather than uniform refinement.

Second, we propose a \emph{frozen-remeshing-branch} strategy for approximately differentiable parameter calibration with topology changes. A forward calculation at the current parameter anchor generates a remeshing branch---the mesh sequence and associated transfer operators along the loading path---which is frozen inside a parameter trust region. Automatic differentiation can then propagate through the CPFEM solves and fixed transfers on that branch, while the branch is regenerated when the accepted parameters leave the trusted neighborhood. This piecewise-differentiable construction enables gradient-based calibration against both macroscopic stress--strain data and local lattice-rotation observables without treating discrete mesh construction as smoothly differentiable. The remainder of the paper presents the finite-strain CPFEM formulation, the structure-driven remeshing and frozen-branch methodologies, and numerical demonstrations of remeshing accuracy, large-deformation capability, and inverse-calibration performance.

\section{Crystal Plasticity Simulation Framework}
\label{sec:cpfem_jax}

Firstly, we will give a brief introduction of CPFEM. Let $\Omega_0\subset\mathbb{R}^3$
denote the reference configuration of a polycrystalline RVE. The deformation
map is written as
\begin{equation}
    \mathbf{x}=\boldsymbol{\varphi}(\mathbf{X},t),
    \qquad
    \mathbf{F}=\frac{\partial \boldsymbol{\varphi}}{\partial \mathbf{X}},
\end{equation}
where $\mathbf{F}$ is the deformation gradient. Under quasi-static loading,
the balance equation in the reference configuration is
\begin{equation}
    \mathrm{Div}\,\mathbf{P}=\mathbf{0}
    \quad \mathrm{in}\ \Omega_0,
    \label{eq:balance}
\end{equation}
with displacement and traction boundary conditions imposed on
$\Gamma_D$ and $\Gamma_N$, respectively. After finite element discretization,
Eq.~\eqref{eq:balance} gives the nonlinear residual system
\begin{equation}
    \mathbf{R}(\mathbf{U},\boldsymbol{\xi};\boldsymbol{\theta})=\mathbf{0},
    \label{eq:global_residual}
\end{equation}
where $\mathbf{U}$ is the vector of nodal degrees of freedom,
$\boldsymbol{\xi}$ collects integration-point internal variables, and
$\boldsymbol{\theta}$ denotes material or design parameters.

The crystal plasticity model follows the standard multiplicative
decomposition
\begin{equation}
    \mathbf{F}=\mathbf{F}^{e}\mathbf{F}^{p},
    \label{eq:decomp}
\end{equation}
where $\mathbf{F}^{e}$ represents the elastic lattice distortion and
$\mathbf{F}^{p}$ represents the plastic deformation induced by crystallographic
slip. The elastic strain is defined as
\begin{equation}
    \mathbf{E}^{e}
    =
    \frac{1}{2}
    \left[
    (\mathbf{F}^{e})^{T}\mathbf{F}^{e}-\mathbf{I}
    \right],
\end{equation}
and the second Piola--Kirchhoff stress is
\begin{equation}
    \mathbf{S}=\mathbb{C}:\mathbf{E}^{e},
\end{equation}
where $\mathbb{C}$ is the anisotropic elastic stiffness tensor. The
corresponding Cauchy stress and first Piola--Kirchhoff stress are evaluated
from $\mathbf{F}^{e}$, $\mathbf{S}$, and $\mathbf{F}$ and are then used in the
global finite element residual.

Plastic flow is described by the plastic velocity gradient
\begin{equation}
    \mathbf{L}^{p}
    =
    \dot{\mathbf{F}}^{p}(\mathbf{F}^{p})^{-1}
    =
    \sum_{\alpha=1}^{N_s}
    \dot{\gamma}^{\alpha}
    \mathbf{s}_{0}^{\alpha}\otimes\mathbf{m}_{0}^{\alpha},
    \label{eq:Lp}
\end{equation}
where $N_s$ is the number of slip systems, $\dot{\gamma}^{\alpha}$ is the slip
rate, and $\mathbf{s}_{0}^{\alpha}$ and $\mathbf{m}_{0}^{\alpha}$ are the slip
direction and slip-plane normal in the sample reference frame. The resolved
shear stress is
\begin{equation}
    \tau^{\alpha}
    =
    \mathbf{S}:
    \left(
    \mathbf{s}_{0}^{\alpha}\otimes\mathbf{m}_{0}^{\alpha}
    \right).
\end{equation}
A rate-dependent power law is used for slip activation:
\begin{equation}
    \dot{\gamma}^{\alpha}
    =
    \dot{\gamma}_{0}
    \left|
    \frac{\tau^{\alpha}}{g^{\alpha}}
    \right|^{1/m}
    \mathrm{sign}(\tau^{\alpha}),
    \label{eq:slip_rate}
\end{equation}
where $\dot{\gamma}_{0}$ is the reference slip rate, $m$ is the rate-sensitivity
exponent, and $g^{\alpha}$ is the slip resistance. The evolution of slip
resistance is written as
\begin{equation}
    \dot{g}^{\alpha}
    =
    \sum_{\beta=1}^{N_s}
    h^{\alpha\beta}
    \left|
    \dot{\gamma}^{\beta}
    \right|,
    \label{eq:gdot}
\end{equation}
with the hardening interaction
\begin{equation}
    h^{\alpha\beta}
    =
    q^{\alpha\beta} h_0
    \left|
    1-\frac{g^{\beta}}{g_{\mathrm{sat}}}
    \right|^{a}
    \mathrm{sign}
    \left(
    1-\frac{g^{\beta}}{g_{\mathrm{sat}}}
    \right).
    \label{eq:hardening}
\end{equation}
Here $h_0$, $g_{\mathrm{sat}}$, and $a$ are hardening parameters, and
$q^{\alpha\beta}$ distinguishes self and latent hardening.

At each load increment, the constitutive update is solved implicitly at every
integration point. In the present implementation, the stored history variables
include the inverse plastic deformation gradient, slip resistance, accumulated
slip, and crystallographic orientation,
\begin{equation}
    \boldsymbol{\xi}
    =
    \left\{
    \mathbf{F}^{p^{-1}},
    \{g^{\alpha}\},
    \{\gamma^{\alpha}\},
    \mathbf{Q}
    \right\}.
    \label{eq:internal_variables}
\end{equation}
The inverse plastic deformation gradient is updated in a backward-Euler form,
\begin{equation}
    \left(\mathbf{F}^{p}_{n+1}\right)^{-1}
    =
    \left(\mathbf{F}^{p}_{n}\right)^{-1}
    \left(
    \mathbf{I}
    -
    \Delta t\,\mathbf{L}^{p}_{n+1}
    \right).
    \label{eq:Fpinv_update}
\end{equation}
Given $\mathbf{F}_{n+1}$ and the previous internal variables
$\boldsymbol{\xi}_{n}$, the local stress and updated state are obtained from a
nonlinear constitutive residual,
\begin{equation}
    \mathbf{R}_{c}
    \left(
    \mathbf{S}_{n+1},
    \mathbf{F}_{n+1},
    \boldsymbol{\xi}_{n}
    \right)
    =
    \mathbf{0}.
    \label{eq:constitutive_residual}
\end{equation}
This local problem couples the elastic law, slip kinetics, hardening law, and
plastic update. The elastic deformation gradient also provides the lattice
rotation through polar decomposition,
\begin{equation}
    \mathbf{F}^{e}=\mathbf{R}^{e}\mathbf{U}^{e},
\end{equation}
from which a scalar rotation angle can be defined as
\begin{equation}
    \omega
    =
    \cos^{-1}
    \left(
    \frac{\mathrm{tr}(\mathbf{R}^{e})-1}{2}
    \right).
    \label{eq:rotation_angle}
\end{equation}
This local rotation measure will later be used to define the grain-boundary
rotation objective in the inverse correction problem. 

The global equilibrium problem is solved by an outer Newton iteration over
the nodal displacement vector. At load increment $n+1$, the trial
displacement $\mathbf{U}^{(k)}_{n+1}$ determines the deformation gradient
$\mathbf{F}^{(k)}_{n+1}$ at every quadrature point. The local constitutive
problem in Eq.~\eqref{eq:constitutive_residual} is then solved at each
quadrature point to obtain the first Piola--Kirchhoff stress
$\mathbf{P}^{(k)}_{n+1}$ and the consistent material tangent
$\mathbb{A}^{(k)}_{n+1}=\partial \mathbf{P}_{n+1}/\partial
\mathbf{F}_{n+1}$. These quantities define the assembled finite element
residual
\begin{equation}
    \mathbf{R}^{(k)}
    =
    \mathbf{f}_{\mathrm{int}}
    \left(
    \mathbf{U}^{(k)}_{n+1},
    \boldsymbol{\xi}_{n}
    \right)
    -
    \mathbf{f}_{\mathrm{ext},n+1},
    \label{eq:global_newton_residual}
\end{equation}
where $\mathbf{f}_{\mathrm{int}}$ is obtained from the weak form of
Eq.~\eqref{eq:balance} using the quadrature stresses. The Newton correction is
computed from
\begin{equation}
    \mathbf{K}^{(k)} \Delta \mathbf{U}^{(k)}
    =
    -
    \mathbf{R}^{(k)},
    \qquad
    \mathbf{K}^{(k)}
    =
    \frac{\partial \mathbf{R}}{\partial \mathbf{U}}
    \left(
    \mathbf{U}^{(k)}_{n+1}
    \right),
    \label{eq:global_newton_linearization}
\end{equation}
and the displacement is updated as
\begin{equation}
    \mathbf{U}^{(k+1)}_{n+1}
    =
    \mathbf{U}^{(k)}_{n+1}
    +
    \eta^{(k)} \Delta \mathbf{U}^{(k)} ,
    \label{eq:global_newton_update}
\end{equation}
where $\eta^{(k)}$ is a line-search parameter used to improve robustness for
the strongly nonlinear crystal-plasticity response. After the residual norm
and displacement correction satisfy the prescribed convergence tolerances, the
solution $\mathbf{U}_{n+1}$ is accepted and the quadrature-point history
variables are committed. Thus, the calculation has a nested structure: the
outer Newton iteration enforces global force balance, while the inner
quadrature-level Newton iteration enforces the implicit crystal-plasticity
constitutive update. In JAX-CPFEM, both the quadrature-level tangent and the
global linearization are evaluated through automatic differentiation and
vectorized array operations, which avoids manual derivation of
case-specific Jacobians and enables efficient GPU execution
\citep{hu2025efficient}.

In conventional CPFEM, grain boundaries are usually represented implicitly through an abrupt change of crystallographic orientation between neighboring grains. The displacement field remains continuous across the grain boundary when conforming finite element meshes are used, while the deformation gradient, stress, strain, slip activity and internal variables are allowed to be discontinuous across element boundaries. Therefore, a special discontinuity treatment is not required unless grain-boundary-specific mechanisms, such as grain boundary sliding, decohesion, or dislocation transmission, are explicitly modeled. In the present framework, grain boundaries are assumed to be perfectly bonded internal interfaces. Their mechanical influence is captured through the compatibility of deformation and traction equilibrium between adjacent grains with different crystallographic orientations.

JAX-CPFEM provides the differentiable and GPU-accelerated implementation of
the above CPFEM formulation. Its main advantage for the present work is that
the local constitutive tangent and the sensitivity of the global response can be
obtained through automatic differentiation rather than manually derived
case-specific Jacobians. In addition, element-wise and quadrature-point-wise
operations are vectorized in an array-programming style, which enables efficient
batched evaluation on GPUs. Therefore, once the forward CPFEM problem is
written as a differentiable computational graph, gradients of a reduced objective
with respect to material or microstructural parameters can be evaluated and used
by gradient-based optimizers.

In this work, JAX-CPFEM serves as the differentiable mechanical core. The
following sections extend this fixed-mesh solver to a two-stage large-deformation
workflow with structure-driven remeshing. The remeshing step adaptively
refines regions near grain boundaries and regions with large internal-variable
activity. During the backward pass, the remeshed topology and transfer pattern
are frozen, so that gradients are propagated through the CPFEM solves and the
fixed state-transfer operator while derivatives associated with mesh-topology
changes are omitted.

\section{Methodology}
\label{sec:methodology}

\begin{figure}[htbp]
    \centering
    \includegraphics[scale=0.48]{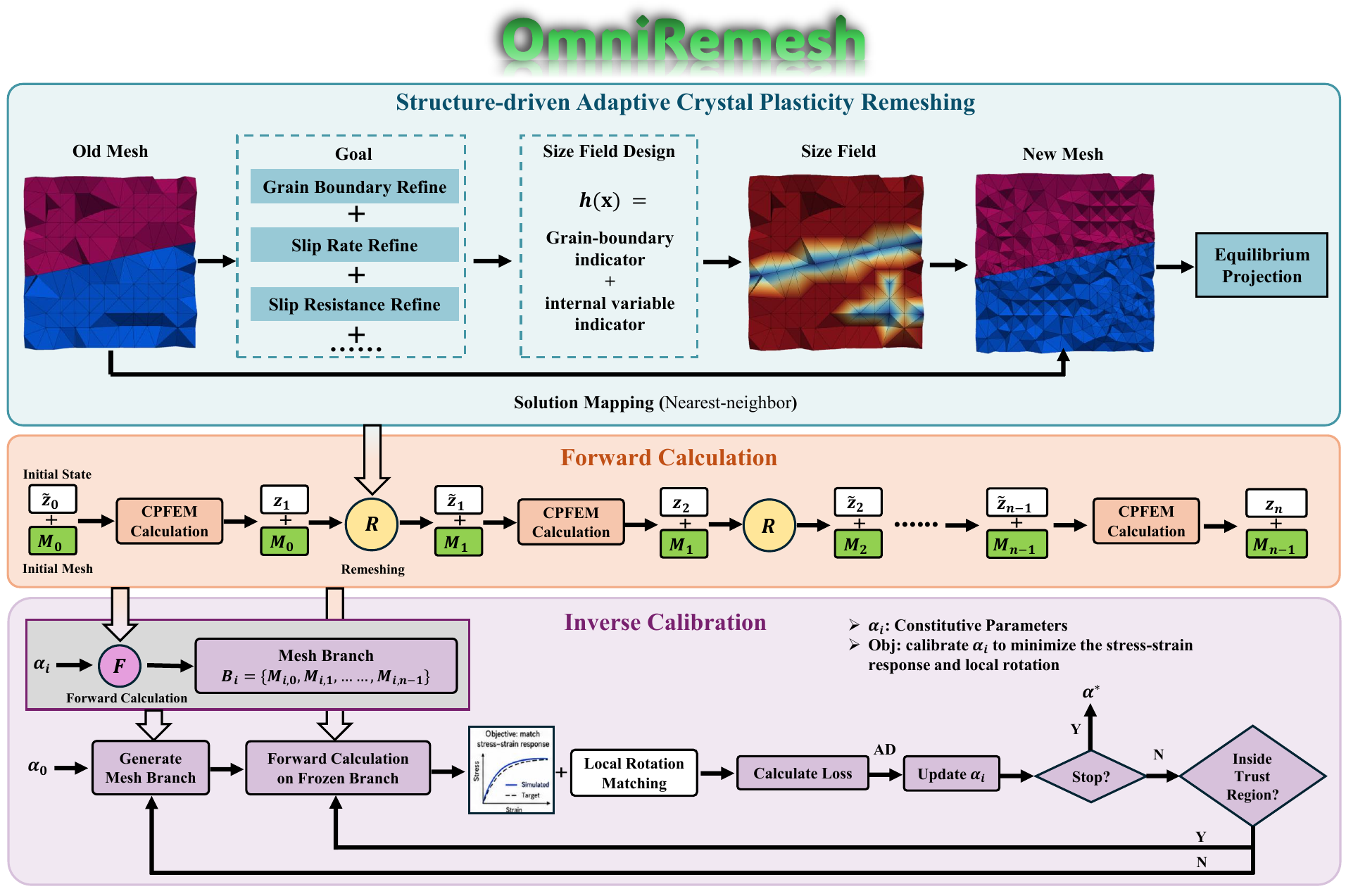}
    \caption{Overview of OmniRemesh. The first part constructs a size field
    from grain-boundary and CPFEM internal-variable indicators, the middle block
    advances the remeshed forward CPFEM calculation, and the lower block uses
    frozen remeshing branches for inverse calibration.}
    \label{overview}
\end{figure}

The overview of OmniRemesh is organized into three connected components, as shown in Fig.~\ref{overview}.
The first part introduces the crystal plasticity adaptive remeshing module, including generating the size field from the structure information of the old solution state, generating the new mesh, and equilibrium projection. The second part illustrates the forward CPFEM remeshing calculation. The final part performs the inverse calibration of constitutive parameters according to the target stress-strain response and target local rotation using remeshing calculation and the approximated differentiable property, which is implemented by the frozen remeshing branches inside the trust region.

\subsection{Structure-driven crystal plasticity remeshing}
\label{subsec:sf_remeshing}
The core idea of adaptive meshing for crystal plasticity is performing mesh refinement near the grain boundary and the region where internal variables get large (hot spots), so the CPFEM calculation will capture more local details and overcome mesh distortion. Specifically, we first define the size field on the current state $\mathbf{x}_i$, then generate the new mesh and transfer the internal variables from the old mesh to the new mesh. Finally, an equilibrium projection step is required to guarantee the equilibrium.

To define the size field, a characteristic length is defined by
\begin{equation}
    L_c = \min_{j=1,2,3}\left(x_j^{\max}-x_j^{\min}\right),
    \label{eq:method_characteristic_length}
\end{equation}
where $x_j^{\max}$ and $x_j^{\min}$ are the maximum and minimum deformed
coordinates in the Cartesian direction $j$. Then we define
\begin{equation}
    h_{\mathrm{bg}}=c_{\mathrm{bg}}L_c,
    \qquad
    h_{\mathrm{gb}}^{\min}=c_{\mathrm{gb}}L_c,
    \qquad
    h_{\mathrm{hot}}^{\min}=c_{\mathrm{hot}}L_c,
    \qquad
    r_a=\eta_a L_c,\quad a\in\{\mathrm{gb},\mathrm{hot}\},
    \label{eq:method_size_scales_simple}
\end{equation}
where $h_{\mathrm{bg}}$ is the coarse background mesh size,
$h_{\mathrm{gb}}^{\min}$ is the target size near grain boundaries,
$h_{\mathrm{hot}}^{\min}$ is the target size near internal-variable hot spots,
and $r_a$ is the transition radius for source type $a$. The dimensionless constants
$c_{\mathrm{bg}}$, $c_{\mathrm{gb}}$, $c_{\mathrm{hot}}$, and $\eta_a$ control
relative mesh resolution. 

Then consider the mesh refinement near the grain-boundary. Let $\Gamma_{\mathrm{gb}}$ be
the union of triangular faces shared by neighboring tetrahedra with different
grain identifiers. For a node at $\mathbf{x}$, define its distance to the grain
boundary by
\begin{equation}
    d_{\mathrm{gb}}(\mathbf{x})
    =
    \min_{\mathbf{y}\in\Gamma_{\mathrm{gb}}}
    \|\mathbf{x}-\mathbf{y}\|_2,
    \label{eq:method_gb_distance}
\end{equation}
where $\|\cdot\|_2$ is the Euclidean norm. The pure grain-boundary driven size field is
\begin{equation}
    h_{\mathrm{gb}}(\mathbf{x})
    =
    h_{\mathrm{gb}}^{\min}
    +
    \left(h_{\mathrm{bg}}-h_{\mathrm{gb}}^{\min}\right)
    \min\left(\frac{d_{\mathrm{gb}}(\mathbf{x})}{r_{\mathrm{gb}}},1\right).
    \label{eq:method_gb_size}
\end{equation}
This expression says that nodes on the grain boundary receive the fine size
$h_{\mathrm{gb}}^{\min}$, while nodes farther than $r_{\mathrm{gb}}$ from the boundary
return to the background size $h_{\mathrm{bg}}$. The transition is gradual, so
the mesh does not jump abruptly from fine to coarse elements. 

As for large internal variables hot spots mesh refinement, let $K$ be a tetrahedral element, $q=1,\ldots,n_q$ be the
quadrature-point index, and $\alpha_s=1,\ldots,N_s$ be the slip-system index.
The slip rate and slip resistance are denoted by
$\dot{\gamma}_{Kq}^{\alpha_s}$ and $g_{Kq}^{\alpha_s}$, respectively. The two
raw element indicators are the average slip activity $A_K$ and the average slip
resistance $G_K$,
\begin{equation}
    A_K
    =
    \frac{1}{n_q}\sum_{q=1}^{n_q}
    \left(
    \sum_{\alpha_s=1}^{N_s}
    \left|\dot{\gamma}_{Kq}^{\alpha_s}\right|^2
    \right)^{1/2},
    \qquad
    G_K
    =
    \frac{1}{n_q}\sum_{q=1}^{n_q}
    \max_{1\leq \alpha_s\leq N_s} g_{Kq}^{\alpha_s}.
    \label{eq:method_hotspot_raw}
\end{equation}
Here $A_K$ is large where slip is active, and $G_K$ is large where local
hardening is strong. Because these two quantities have different physical
scales, each is first rescaled to the interval $[0,1]$. Denote these rescaled
fields by $\widehat{A}_K$ and $\widehat{G}_K$. The hot-spot score is 
\begin{equation}
    I_{\mathrm{hot},K}
    =
    \max\left(\widehat{A}_K, \widehat{G}_K\right).
    \label{eq:method_hotspot_indicator}
\end{equation}
Elements with a
large $I_{\mathrm{hot},K}$ form a hot-spot source cloud
$\mathcal{C}_{\mathrm{hot}}$ through their centroids. If
$d_{\mathrm{hot}}(\mathbf{x})$ is the distance from node $\mathbf{x}$ to this
source cloud, the pure hot-spot sriven size field is
\begin{equation}
    h_{\mathrm{hot}}(\mathbf{x})
    =
    h_{\mathrm{hot}}^{\min}
    +
    \left(h_{\mathrm{bg}}-h_{\mathrm{hot}}^{\min}\right)
    \min\left(\frac{d_{\mathrm{hot}}(\mathbf{x})}{r_{\mathrm{hot}}},1\right).
    \label{eq:method_hotspot_size}
\end{equation}
Thus the mesh is refined around high-activity or high-hardening regions and
returns smoothly to the background size away from them.

The final nodal size field combines the background, grain-boundary, and hot-spot
requests by taking the smallest requested size,
\begin{equation}
    h(\mathbf{x}_i)
    =
    \min\left\{
    h_{\mathrm{bg}},
    h_{\mathrm{gb}}(\mathbf{x}_i),
    h_{\mathrm{hot}}(\mathbf{x}_i)
    \right\}.
    \label{eq:method_size_field}
\end{equation}
We can generate the new mesh $\mathcal{M}^{\mathrm{new}}$ according to the size field on the current state, and perform the internal variables transfer using nearest-neighbor interpolation. However, the new state may deviate from equilibrium; thus, an extra projection solving step is introduced to enforce the balance condition in Eq. \ref{eq:balance}, after which we get the new solution state $\mathbf{\tilde{\mathbf{z}}}$. 

\subsection{Forward remeshing calculation}
\label{subsec:forward_calculation}

The forward remeshing calculation is essentially several CPFEM calculations along the prescribed loading path coupled with regular remeshing procedures, as shown in the middle block of Fig.~\ref{overview}. Starting from the initial state $\mathbf{z}_0$ and initial mesh $\mathcal{M}_0$, the entire loading process is evenly divided into $n$ parts and perform $n$ times corresponding CPFEM calculations. After each CPFEM calculation, a complete remeshing procedure $\mathcal{R}$ will be followed, including generating adaptive mesh from the size field, mapping the internal variables, and perform equilibrium projection on the new mesh. The forward remeshing calculation can be formulated by 
\begin{equation}
\begin{aligned}
    \mathbf{z}_{j+1}, \mathcal{M}_{j}
    &= \mathcal{S}_{j}\left(
        \tilde{\mathbf{z}}_{j},
        \mathcal{M}_{j};
        \boldsymbol{\alpha}
    \right), \\
    \tilde{\mathbf{z}}_{j+1}, \mathcal{M}_{j+1}
    &= \mathcal{R}_j\left(
        \mathcal{M}_{j},
        \mathbf{z}_{j+1};
        \boldsymbol{\alpha}
    \right), \quad
    j = 0, 1, \dots, n-1.
    \label{eq:method_forward_flow}
\end{aligned}
\end{equation}
where $\mathcal{M}_j$ is the mesh at remeshing stage $j$, and $\mathbf{z}_j$ is the corresponding CPFEM state. The state $\mathbf{z}_j$
contains the nodal displacement, the quadrature-point internal variables in
Eq.~\eqref{eq:internal_variables}, and the stress or rotation quantities
extracted from the CPFEM solve. Let $\boldsymbol{\alpha}$ denote the
constitutive parameter vector used in the material update. On a fixed mesh, the
CPFEM loading calculation from one remeshing stage to the next is denoted by
$\mathcal{S}_j$.

\subsection{Inverse calibration with frozen remeshing branches}
\label{subsec:opt_frozen_remesh_calibration}

We want to further implement our remeshing calculation on inverse calibration of constitutive parameters $\boldsymbol{\alpha}$. The goal is that two responses match the target:
the macroscopic stress--strain curve and the local lattice rotation near the
selected grain-boundary regions. If the calculation were performed on a fixed
mesh, the parameter correction could be performed directly by the automatic differentiation of 
JAX-CPFEM \cite{hu2026efficient}. With adaptive
remeshing, however, changing $\boldsymbol{\alpha}$ can change the size field and
therefore the topology of the remeshed sequence. The derivative of this discrete
mesh-topology change is not available in a useful smooth form. We therefore use
a parameter-trust strategy: around the current accepted parameter, a forward
remeshing calculation generates a frozen remeshing branch; inside the trust
region, optimization reuses this branch and applies automatic differentiation
only through the CPFEM calculations on the fixed branch. When the accepted
parameter leaves the trusted range, the remeshing branch is regenerated.

Let $\bar{\boldsymbol{\alpha}}_k$ be the accepted parameter at which branch $k$
is generated. A full forward remeshing calculation is first run at
$\bar{\boldsymbol{\alpha}}_k$ using the procedure in
Eq.~\eqref{eq:method_forward_flow}. This produces a sequence of remeshed meshes
$\mathcal{M}_{k,0},\mathcal{M}_{k,1},\ldots,\mathcal{M}_{k,n}$, where
$\mathcal{M}_{k,j}$ is the mesh used after the $j$-th remeshing stage. The
frozen branch is written as
\begin{equation}
    \mathcal{B}_k
    =
    \left\{
    \mathcal{M}_{k,0},\mathcal{M}_{k,1},\ldots,\mathcal{M}_{k,n}
    \right\}.
    \label{eq:method_branch_definition}
\end{equation}
For a trial parameter $\boldsymbol{\alpha}$ inside the current trust region, the
forward response is evaluated on the frozen branch. The loss is the sum of a normalized
stress--strain mismatch and a normalized local-rotation mismatch. The objective
on frozen branch $k$ is
\begin{equation}
    J_k(\boldsymbol{\alpha})
    =
    c_J\left[
    J_{\sigma}^{(k)}(\boldsymbol{\alpha})
    +\lambda_{\omega}J_{\omega}^{(k)}(\boldsymbol{\alpha})
    \right],
    \label{eq:method_frozen_objective}
\end{equation}
where $J_{\sigma}^{(k)}$ measures the stress--strain mismatch,
$J_{\omega}^{(k)}$ measures the grain-boundary rotation mismatch,
$\lambda_{\omega}$ balances their relative contributions, $c_J$ is a numerical
scaling factor. Their
problem-specific definitions are given in
Sec.~\ref{subsec:results_calibration}. Within $\mathcal{A}_k$, the gradient is
approximated by
\begin{equation}
    \widetilde{\nabla}_{\boldsymbol{\alpha}}J_k
    =
    \left.
    \frac{\partial J(\boldsymbol{\alpha};\mathcal{B})}
         {\partial\boldsymbol{\alpha}}
    \right|_{\mathcal{B}=\mathcal{B}_k},
    \label{eq:method_frozen_gradient}
\end{equation}
which retains the parameter dependence of the CPFEM states but omits the
derivative of the branch construction itself.

The optimization
inside branch $k$ is therefore
\begin{equation}
    \boldsymbol{\alpha}_{k}^{\mathrm{new}}
    =
    \arg\min_{\boldsymbol{\alpha}\in\mathcal{A}_k}
    J_k(\boldsymbol{\alpha}),
    \label{eq:method_branch_optimization}
\end{equation}
where $\mathcal{A}_k$ is the trust region. The gradient used in this
minimization is computed by automatic differentiation through the CPFEM solves
on the frozen branch. The dependence of the mesh sequence
$\mathcal{B}_k$ on $\boldsymbol{\alpha}$ is ignored within this local problem.

The trust region restricts how far the trial parameter can move away from the
branch anchor. With global bounds $\boldsymbol{\alpha}^{\min}$ and
$\boldsymbol{\alpha}^{\max}$ and component-wise trust radii
$\boldsymbol{\Delta}$, it is defined as
\begin{equation}
    \mathcal{A}_k
    =
    \left\{
    \boldsymbol{\alpha}:
    \boldsymbol{\alpha}^{\min}
    \leq
    \boldsymbol{\alpha}
    \leq
    \boldsymbol{\alpha}^{\max},
    \quad
    d_k(\boldsymbol{\alpha})\leq 1
    \right\},
    \qquad
    d_k(\boldsymbol{\alpha})
    =
    \max_m
    \frac{|\alpha_m-\bar{\alpha}_{k,m}|}{\Delta_m}.
    \label{eq:method_trust_region}
\end{equation}
Here $\alpha_m$ and $\bar{\alpha}_{k,m}$ are the $m$-th components of the trial
parameter and branch anchor, and $\Delta_m$ is the allowed movement in that
component. After each accepted update, the distance
$d_k(\boldsymbol{\alpha}_{k}^{\mathrm{new}})$ is checked. If the accepted
parameter remains sufficiently inside the trusted range, the next optimization
batch reuses the same frozen branch. If it approaches or exceeds the boundary,
the accepted parameter becomes the new anchor and a new forward remeshing
calculation is performed to generate a refreshed branch. This procedure gives a
piecewise differentiable calibration scheme: within each trust region, the mesh
branch is fixed and JAX-CPFEM provides the parameter gradient; across trust
regions, the branch is updated to remain consistent with the current
constitutive parameters. The whole algorithm is shown in Algorithm \ref{alg:frozen_branch_calibration}.

\begin{algorithm}[htbp]
\caption{Inverse calibration with frozen remeshing branches}
\label{alg:frozen_branch_calibration}
\begin{algorithmic}[1]
\Require Initial parameter $\boldsymbol{\alpha}_0$, target stress curve
$\{(\varepsilon_i,\widehat{\sigma}_i)\}_{i=1}^{N_\varepsilon}$, target local
rotations $\{\widehat{\omega}_r\}_{r=1}^{N_r}$, trust region $\mathcal{A}_0$
\State Set the accepted parameter
$\boldsymbol{\alpha}_{\mathrm{acc}}\gets\boldsymbol{\alpha}_0$
\State Set branch index $k\gets 0$ and generate initial remesh branch $\mathcal{B}_0$
\While{the stopping criterion is not satisfied}
    \State Forward calculation on frozen remeshing branch
    $\boldsymbol{\alpha}\in\mathcal{A}_k$
    \State Compute the loss $J_k(\boldsymbol{\alpha})$ from Eq.~\eqref{eq:method_frozen_objective}
    \State Minimize $J_k(\boldsymbol{\alpha})$ using AD gradients on the frozen branch
    \State Accept the optimized parameter as
    $\boldsymbol{\alpha}_{\mathrm{acc}}$
    \State Compute $d_k(\boldsymbol{\alpha}_{\mathrm{acc}})$ from
    Eq.~\eqref{eq:method_trust_region}
    \If{$\boldsymbol{\alpha}_{\mathrm{acc}}$ is outside the trust-region}
        \State Set $k\gets k+1$ 
        \State Set the branch anchor
        $\bar{\boldsymbol{\alpha}}_k\gets\boldsymbol{\alpha}_{\mathrm{acc}}$
        \State Run a forward remeshing calculation at
        $\bar{\boldsymbol{\alpha}}_k$ using Eq.~\eqref{eq:method_forward_flow}
        \State Store the frozen branch $\mathcal{B}_k$ according to
        Eq.~\eqref{eq:method_branch_definition}
        \State Define the trust region $\mathcal{A}_k$ by
        Eq.~\eqref{eq:method_trust_region} 
    \EndIf
\EndWhile
\State \Return calibrated parameter
$\boldsymbol{\alpha}^{\star}=\boldsymbol{\alpha}_{\mathrm{acc}}$
\end{algorithmic}
\end{algorithm}

\section{Simulation Results}
In this section, we use 4 examples, the remeshing demo, bicrystal stretching quantitative analysis, multi-grains large stretching qualitative analysis, and inverse calibration, to validate the proposed
OmniRemesh. For all examples, we consider the polycrystalline 304 steel (FCC) RVE with 0.016 mm per side \cite{tran2023asynchronous}.

\subsection{Remeshing demo}
\begin{figure}[htbp]
    \centering
    \includegraphics[scale=0.13]{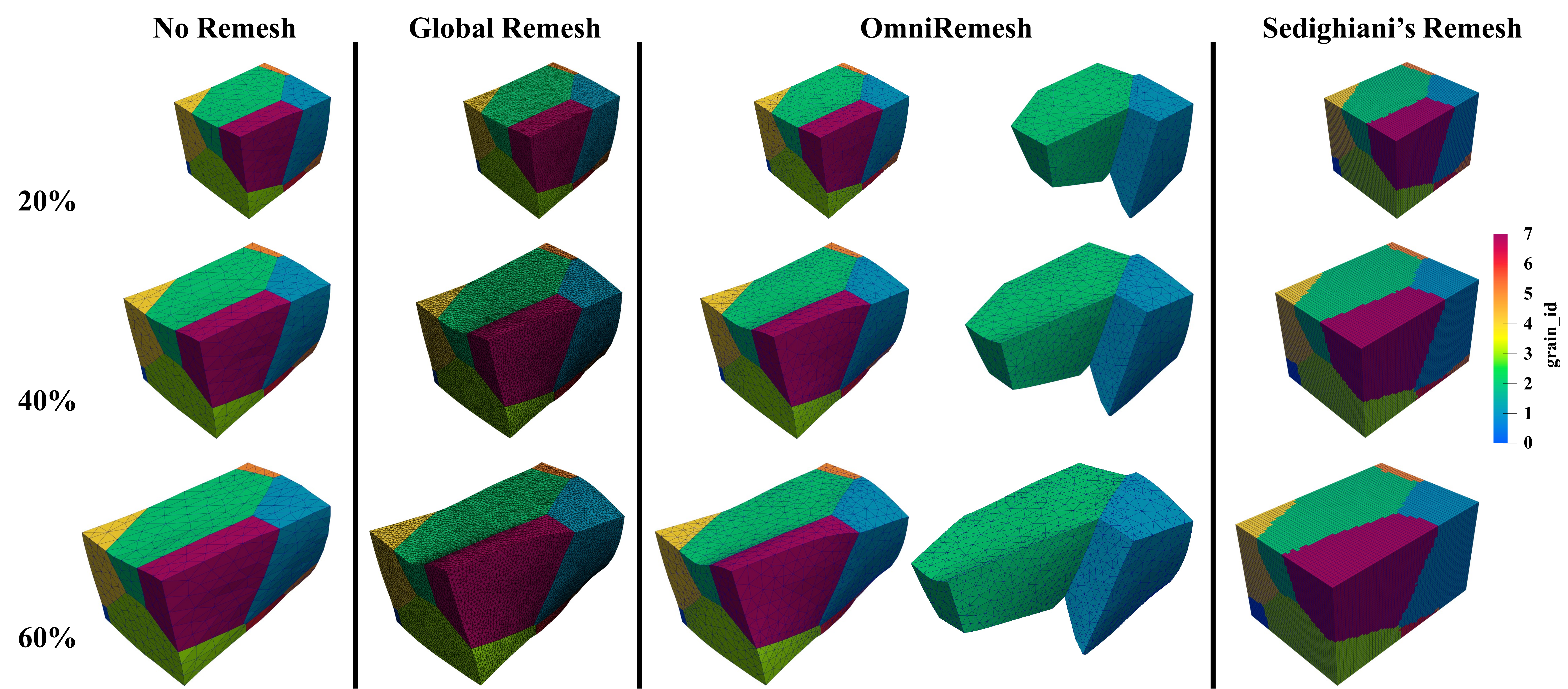}  
    \caption{Remeshing demo of 8-grains RVE}
    \label{remesh demo}
\end{figure}

In this case, we preliminarily illustrate our remeshing ability through an RVE stretching problem. Consider an RVE with 8 grains, and a 60\% tensile loading is prescribed through a displacement boundary condition. The initial mesh is generated through Neper with the relative characteristic lengths (-rcl) 0.5 and tetrahedral elements. The remeshing procedure is performed every 20\% stretching. For each remeshing, the generated size field is fed into mmgpy to generate the new mesh under the current state. We set mesh resolution control parameters as $c_{bg}=0.1$, $c_{gb}=c_{hot}=0.03$, so we can obtain a resolution improvement of nearly 33 times over the old mesh when approaching the grain boundary and large internal variables region. Fig. \ref{remesh demo} shows the RVE mesh with and without remeshing under 20\%, 40\% and 60\% stretching. Compared with the calculation without remeshing and global remeshing, our OmniRemesh results can obtain an adaptive refined mesh for each deformation phase, which is deformed geometry-preserving on the current state. Besides, the first 2 grains meshes are given in the fourth column of Fig. \ref{remesh demo}, illustrating the mesh adaptivity and grain-preserving of our method obviously. Finally, we also implement Sedighiani's \cite{sedighiani2021large} remeshing method, which only supports generating a new structured mesh on the rectangular geometric approximation of the current state, and it is designed only for periodic boundary conditions. Compare with Sedighiani's method, OmniRemesh is more flexible in the choice of element type (almost all element types are acceptable) and boundary condition (displacement and periodic), and supports adaptive local refinement. The characteristics can facilitate the further implementation of OmniRemesh on forward calculation and parameter calibration.

\subsection{Bicrystal Stretching}
\begin{figure}[htbp]
    \centering
    \includegraphics[scale=0.16]{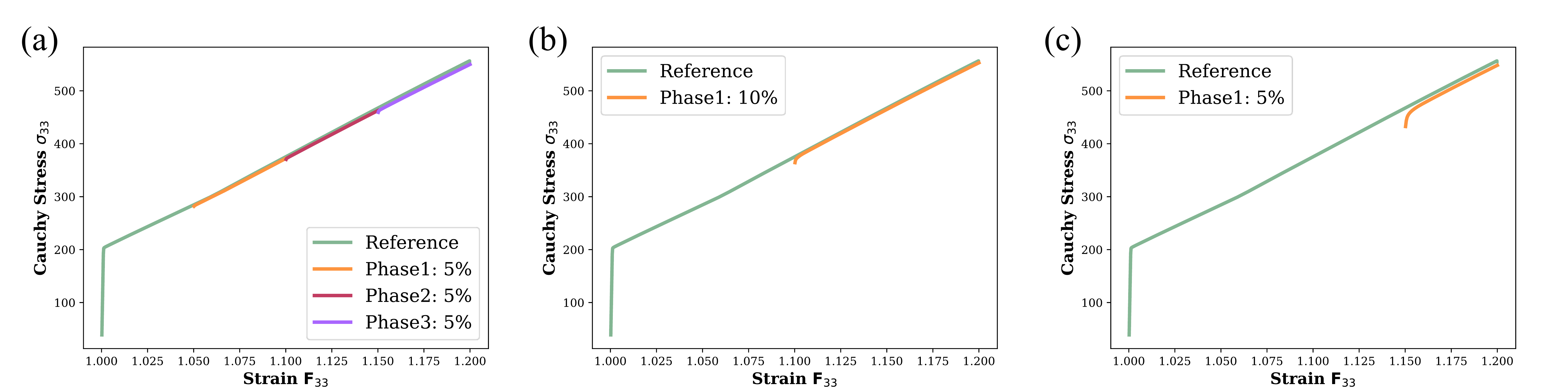}  
    \caption{Different remeshing strategies strain-stress curves. (a) strategy 1, (b) strategy 2, (c) strategy 3.}
    \label{EX2_stress_strain}
\end{figure}

\begin{figure}[htbp]
    \centering
    \includegraphics[scale=0.4]{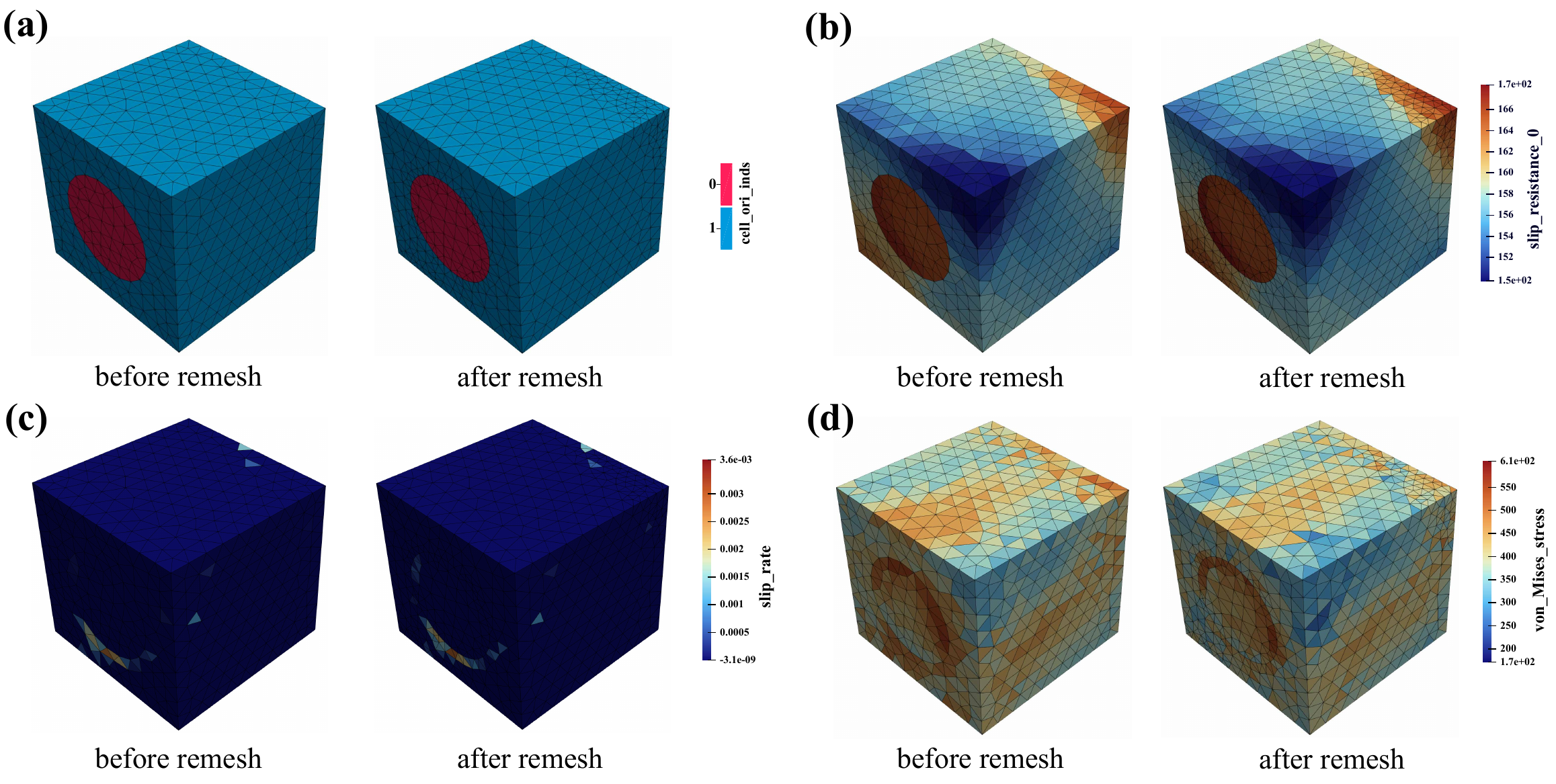}  
    \caption{RVE state before and after remeshing under 10\% stretching in strategy 2. (a) mesh state, (b) slip resistance of 0th slip system, (c) slip rate, (d) Vom-Mises stress.}
    \label{EX2_accuracy_10}
\end{figure}

In this example, the accuracy of OmniRemesh will be verified. Consider a bicrystal RVE, a cube with a cylinder inside, as shown in Fig. \ref{EX2_accuracy_10} (a). Given the displacement boundary condition of 20\% thickness stretching, 4 strategies are used to test the forward remeshing calculation accuracy:
\begin{itemize}
    \item 1 step of 20\% without remeshing (reference)
    \item 4 steps of 5\% (strategy 1)
    \item 2 steps of 10\% (strategy 2)
    \item 2 steps of 15\% + 5\% (strategy 3)
\end{itemize}
The mesh resolution parameters are set the same for all strategies. The stress-strain curves of different remeshing strategies are shonw in Fig. \ref{EX2_stress_strain}, and all these 3 strategies' results are in good agreement with the reference (the results from a high-resolution mesh). The equilibrium projection inherent in the remeshing procedure leads to a minor loss of stress, and it will maintain the original stress state after an additional recovery strain. Furthermore, the recovery strain increases with the remeshing strain.

\begin{figure}[htbp]
    \centering
    \includegraphics[scale=0.35]{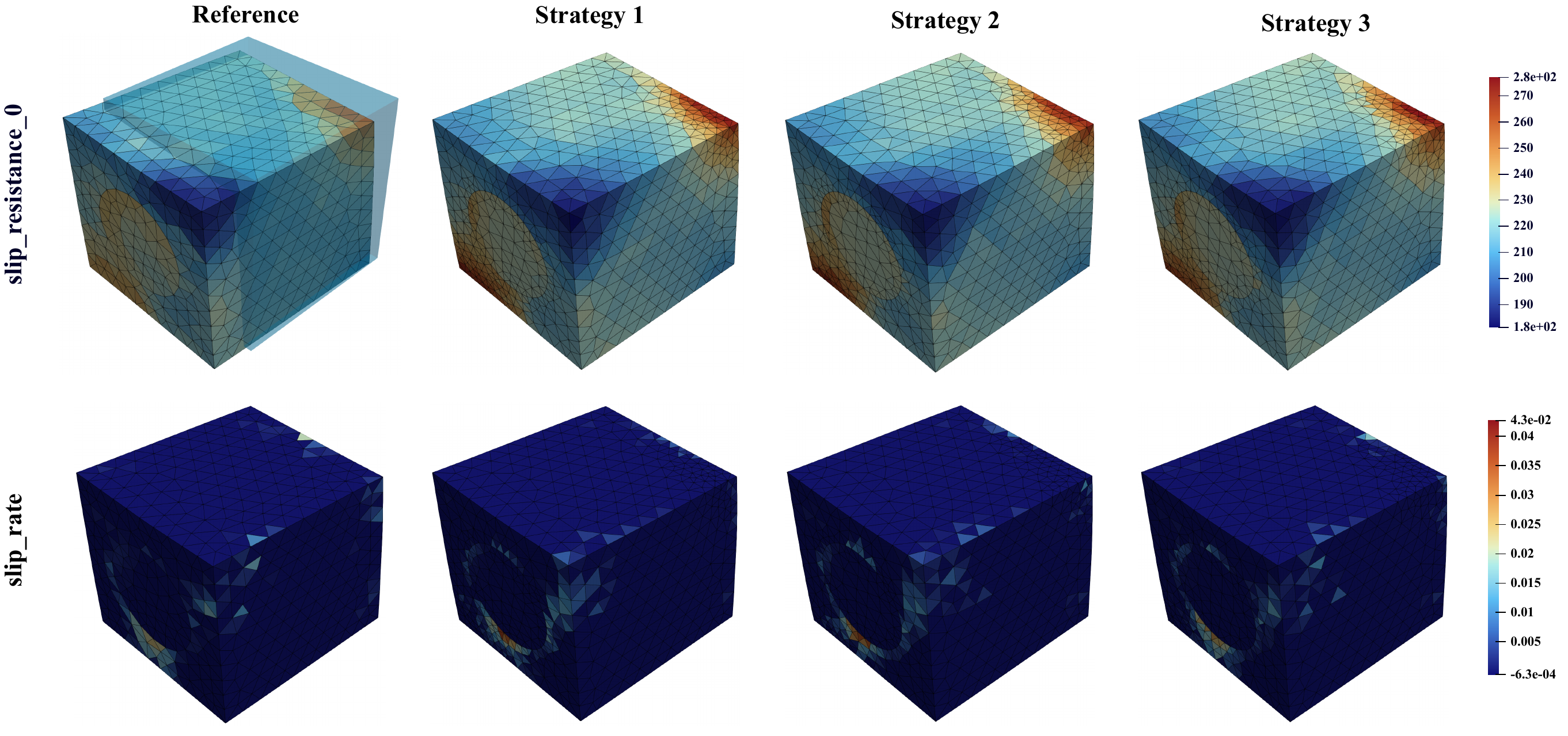}  
    \caption{Different remeshing strategies results}
    \label{EX2_different_strategies}
\end{figure}

In order to illustrate the RVE state changes during the remeshing procedure, we take strategy 2 as an example and give the mesh, slip resistance of 0th slip system, slip rate, and Vom-Mises stress state before and after remeshing, as shown in Fig. \ref{EX2_accuracy_10}. The remeshed internal variables and stress state show good consistency with those before remeshing. Besides, we use the grain-perserving interpolation to further reduce the calculation error from the incorrect information from other grains. Fig. \ref{EX2_different_strategies} shows the final RVE internal variables state of the reference and three strategies calculation, which are all close to each other. This example quantitatively validates the accuracy of the forward calculation of OmniRemesh. 

\subsection{Multi-grains large stretching}
\begin{figure}[htbp]
    \centering
\includegraphics[width=1.0\columnwidth]{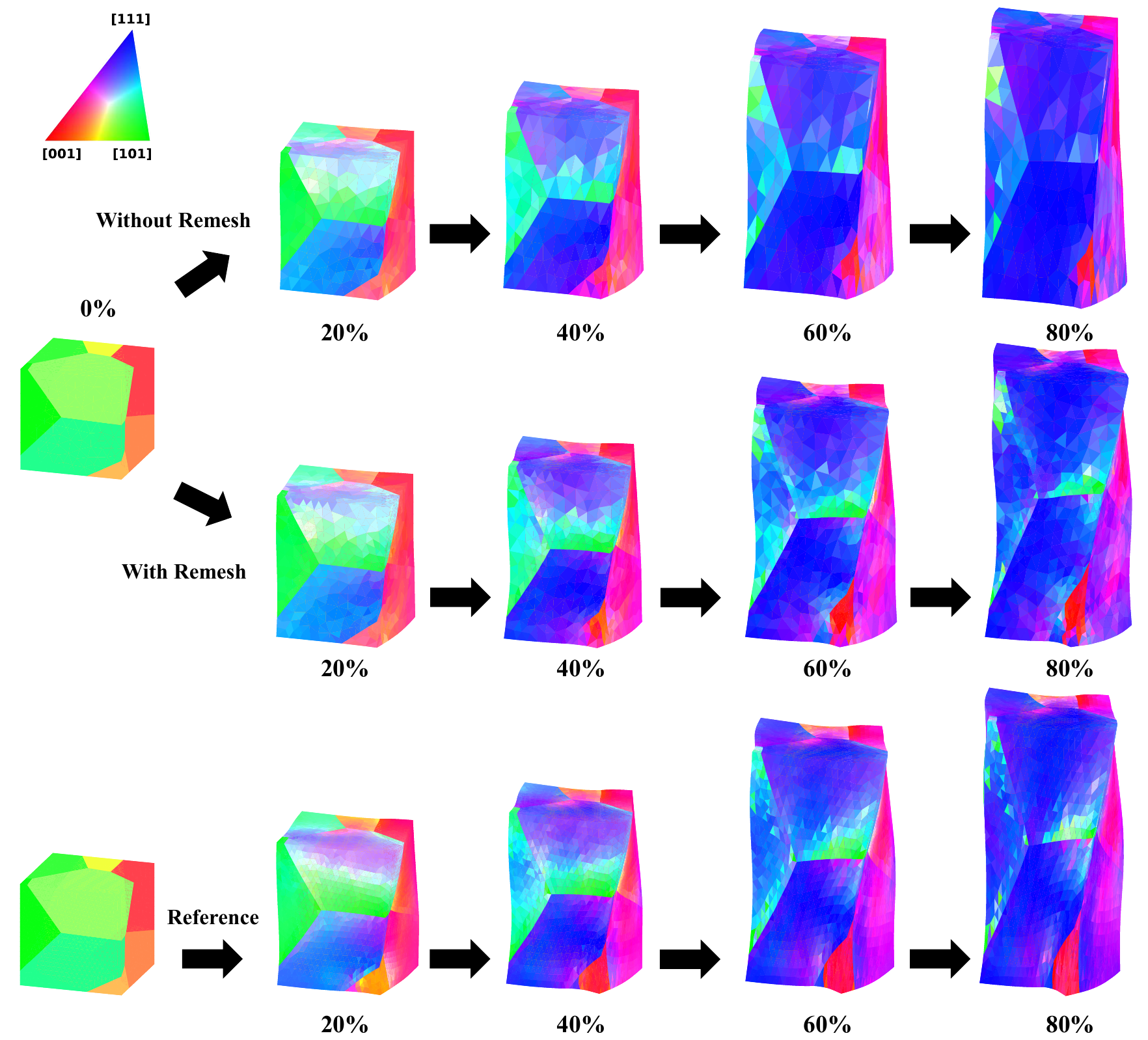}  
    \caption{IPF color maps obtained without remeshing, with adaptive remeshing, and from the dense-mesh reference calculation.}
    \label{cha2_EX3_IPF}
\end{figure}

This example considers the large-deformation response of a polycrystalline representative volume element (RVE). A cubic RVE containing eight randomly oriented grains is generated using Neper, as shown in Fig. \ref{cha2_EX3_IPF}, and is subjected to a displacement boundary condition corresponding to 80\% uniaxial stretching. Three calculations are compared. The first calculation uses the original coarse mesh throughout the loading process without remeshing. The second starts from the same mesh but applies the proposed adaptive remeshing procedure after every 20\% stretching, with the resolution parameters set to $c_{bg}=0.1$ and $c_{gb}=c_{hot}=0.05$. These parameters provide higher resolution near grain boundaries and in regions with large internal-variable gradients. The third calculation employs a uniformly dense mesh without remeshing and serves as the reference solution. This reference enables the accuracy of the two coarse-mesh calculations to be evaluated independently of the remeshing procedure.

Fig. \ref{cha2_EX3_IPF} compares the inverse pole figure (IPF) maps at different stretching levels. All three calculations reproduce the overall deformation and grain-orientation evolution. However, as the deformation increases, the coarse mesh without remeshing becomes progressively distorted and cannot resolve the detailed intragranular orientation variations observed in the dense-mesh reference solution. Its IPF distribution is consequently represented by relatively large, nearly uniform patches. The adaptively remeshed result agrees more closely with the reference solution, particularly at 60\% and 80\% stretching, and captures finer orientation gradients and localized features within the grains. This comparison demonstrates that adaptive refinement reduces the loss of local information caused by mesh coarseness and accumulated element distortion.

\begin{figure}[htbp]
    \centering
\includegraphics[width=1.0\columnwidth]{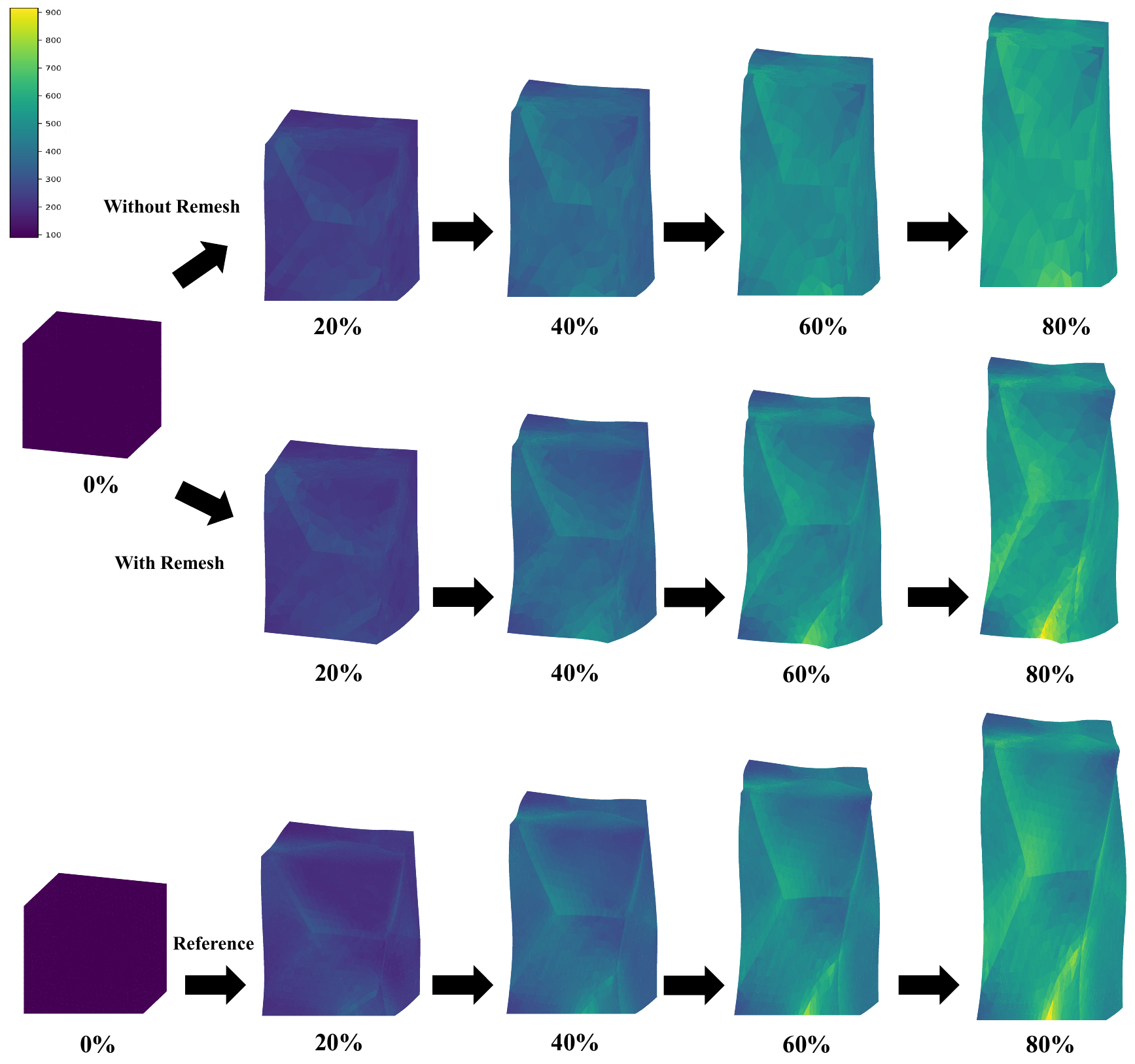}  
    \caption{Slip resistance on the 0th slip system obtained without remeshing, with adaptive remeshing, and from the dense-mesh reference calculation.}
    \label{cha2_EX3_slip_resistance}
\end{figure}

The same improvement is observed in the slip-resistance field shown in Fig. \ref{cha2_EX3_slip_resistance}. The dense-mesh reference solution contains localized hardening patterns and relatively sharp spatial gradients that are smoothed or missed by the calculation without remeshing. By placing finer elements near grain boundaries and regions with large internal-variable gradients, the remeshed calculation recovers more of these local features and produces a field distribution closer to the reference. The agreement becomes especially clear at the larger stretching levels, where the benefit of maintaining local mesh resolution is most pronounced.

\begin{figure}[htbp]
    \centering
\includegraphics[width=0.6\columnwidth]{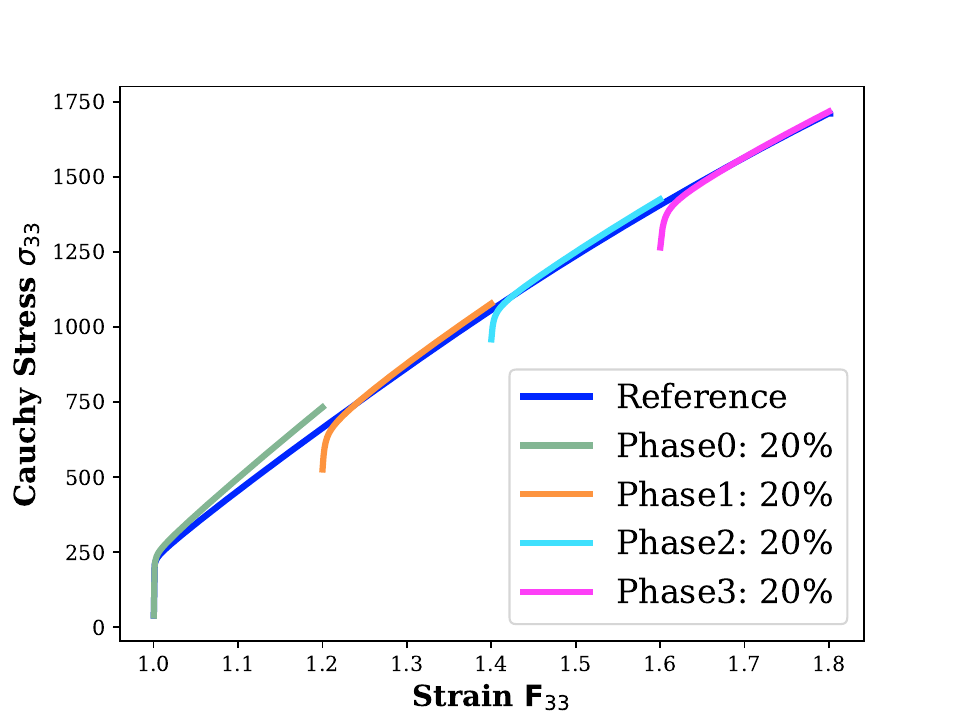}  
    \caption{Stress--strain response during 80\% stretching. The blue curve denotes the dense-mesh reference solution, and the colored segments denote successive 20\% loading phases of the adaptively remeshed calculation.}
    \label{cha2_EX3_stress_strain}
\end{figure}

The macroscopic response provides further evidence of this improvement. Fig. \ref{cha2_EX3_stress_strain} compares the stress--strain response of the remeshed calculation with the dense-mesh reference, shown by the blue curve. The colored segments represent the four successive 20\% loading phases. During the first phase, before any remeshing has been performed, the coarse-mesh response deviates noticeably from the reference. After remeshing, each subsequent phase approaches and follows the reference curve more closely. A short stress-recovery interval appears immediately after each remeshing operation because the transferred state must be re-equilibrated on the new mesh; nevertheless, the discrepancy decreases rapidly as loading continues. The combined macroscopic and local-field comparisons show that the proposed remeshing procedure not only permits the simulation to reach 80\% stretching, but also improves its accuracy relative to the calculation performed on the original coarse mesh by better preserving grain-scale heterogeneity and localized material response.

\subsection{Inverse calibration}
\label{subsec:results_calibration}
\subsubsection{Task setting}
We consider a synthetic inverse-calibration problem for a cubic polycrystalline
RVE containing 50 randomly oriented grains. The specimen is subjected to
displacement-controlled uniaxial tension along the $z$ direction up to an
engineering strain of $10\%$. The loading path is divided into 40 increments,
and one adaptive remeshing operation is performed after the twentieth
increment. A forward remeshing calculation with a prescribed parameter vector
is used to generate the target data. These data comprise the macroscopic axial
stress at all 40 loaded states and the local lattice rotations in 20 selected
grain-boundary regions at the final loading state. Since the reference
parameters are known, this synthetic example provides a controlled test of both
response matching and parameter recovery.

Six spatially uniform, dimensionless correction coefficients are calibrated.
Each coefficient multiplies one nominal constitutive quantity,
\begin{equation}
    p_m(\alpha_m)
    =
    \alpha_m p_m^{\mathrm{nom}},
    \qquad m=1,\ldots,6,
    \label{eq:results_parameter_scaling}
\end{equation}
where $p_m^{\mathrm{nom}}$ is the nominal value and $\alpha_m$ is its correction
coefficient. The calibrated quantities, reference coefficients, initial guesses,
and admissible bounds are listed in Table~\ref{tab:calibration_parameters}. A
common initial value $\alpha_m^{(0)}=1.8$ is used for all six coefficients. This
point lies within the admissible domain but is deliberately different from the
reference vector, providing a nontrivial initialization without supplying the
optimizer with prior knowledge of the solution.

\begin{table}[htbp]
\centering
\small
\caption{Constitutive quantities and correction coefficients used in the
synthetic inverse-calibration example.}
\label{tab:calibration_parameters}
\resizebox{\textwidth}{!}{
\begin{tabular}{c l c c c c}
\hline
Coefficient & Constitutive quantity & $p_m^{\mathrm{nom}}$
& $\alpha_m^{(0)}$ & $\alpha_m^{\mathrm{ref}}$
& Admissible range \\
\hline
$\alpha_1$ & Initial slip resistance $g_0$
& $90\,\mathrm{MPa}$ & 1.8 & 2.5355 & $[0.7,3.0]$ \\
$\alpha_2$ & Hardening exponent $a$
& $8$ & 1.8 & 1.6248 & $[0.7,3.0]$ \\
$\alpha_3$ & Hardening modulus $h_0$
& $392.9772\,\mathrm{MPa}$ & 1.8 & 1.8418 & $[0.7,3.0]$ \\
$\alpha_4$ & Saturation slip resistance $g_{\mathrm{sat}}$
& $7295.1754\,\mathrm{MPa}$ & 1.8 & 0.8286 & $[0.7,3.0]$ \\
$\alpha_5$ & Rate-sensitivity exponent $m$
& $1/120$ & 1.8 & 2.7728 & $[0.7,3.0]$ \\
$\alpha_6$ & Latent-hardening ratio $q$
& $1$ & 1.8 & 1.0968 & $[0.7,2.5]$ \\
\hline
\end{tabular}}
\end{table}

The nominal constitutive quantities are adopted from the polycrystalline 304
steel example reported by Hu et al. \citep{hu2026efficient}. The admissible
intervals are imposed on the dimensionless correction coefficients rather than
directly on the dimensional constitutive quantities. Consequently, a bound
$\alpha_m\in[\alpha_m^{\mathrm{L}},\alpha_m^{\mathrm{U}}]$ corresponds to the
physical interval
$p_m\in[\alpha_m^{\mathrm{L}}p_m^{\mathrm{nom}},
\alpha_m^{\mathrm{U}}p_m^{\mathrm{nom}}]$. This coefficient-based formulation
places parameters with substantially different units and magnitudes on similar
numerical scales, thereby improving the conditioning of the optimization. In
Ref. \citep{hu2026efficient}, the bounds were selected to cover the expected
material-parameter ranges indicated by mechanical data, previous calibration
experience, and literature-reported values. They also prevent the bound-constrained
optimizer from exploring nonphysical or excessively large parameter values. Following
this rationale, a broad interval of $[0.7,3.0]$, corresponding to 70--300\% of
the nominal value, is used for the first five coefficients. A slightly more
conservative upper bound of 2.5 is used for the latent-hardening ratio to limit
excessively strong intersystem hardening and reduce the exploration of extreme,
poorly identifiable parameter combinations. These intervals contain both the
initial guess and the prescribed reference vector while remaining sufficiently
broad for a nontrivial parameter-calibration test.

For this example, the stress component in
Eq.~\eqref{eq:method_frozen_objective} is defined on frozen branch $k$ as
\begin{equation}
    J_{\sigma}^{(k)}(\boldsymbol{\alpha})
    =
    \frac{
    \displaystyle\sum_{i=1}^{N_{\varepsilon}}
    \left[
    \sigma_{zz,i}(\boldsymbol{\alpha};\mathcal{B}_k)
    -\widehat{\sigma}_{zz,i}
    \right]^2
    }{
    \displaystyle\max\left(
    \sum_{i=1}^{N_{\varepsilon}}\widehat{\sigma}_{zz,i}^{\,2},
    \epsilon_{\sigma}
    \right)
    },
    \label{eq:results_stress_loss}
\end{equation}
where $N_{\varepsilon}=40$, $\sigma_{zz,i}$ is the simulated volume-averaged
axial stress at loaded state $i$, $\widehat{\sigma}_{zz,i}$ is the corresponding
target value, and $\epsilon_{\sigma}$ is a small positive constant preventing a
zero denominator.

The local rotation targets are evaluated at the final loading state. At
quadrature point $q$ of element $K$, the scalar lattice rotation angle
$\theta_{Kq}$ is computed from the polar rotation of the elastic deformation
gradient according to Eq.~\eqref{eq:rotation_angle}. The target regions are
selected once from the reference solution as follows. First, quadrature points
whose material coordinates lie within a prescribed distance of a grain
boundary are identified as candidates. Each candidate is associated with the
corresponding grain pair and with the grain on which the point lies. The
candidates are then ranked in descending order of $\theta_{Kq}$, so regions with
strong local lattice rotation are considered first. Starting from the
highest-ranked candidate, a new region is accepted only if its center is
sufficiently separated from previously accepted centers on the same side of the
same grain boundary. This separation criterion removes multiple candidates that
represent the same local hotspot. The procedure continues until $N_r=20$
distinct grain-boundary regions have been obtained.

For region $r$, the material coordinate of the accepted candidate defines its
center $\mathbf{c}_r$, and the selected grain pair identifies the associated
grain boundary. These material-space definitions are stored with the reference
data and reused after every remeshing operation; consequently, the objective
tracks the same physical regions even though the elements and quadrature points
change. Let $\Omega_r^{\mathrm{GB}}$ denote the material patch surrounding
$\mathbf{c}_r$ on the selected side of that grain boundary. Its representative
rotation is defined as the weighted average
\begin{equation}
    \omega_r(\boldsymbol{\alpha};\mathcal{B}_k)
    =
    \frac{
    \displaystyle\sum_{(K,q)\in\Omega_r^{\mathrm{GB}}}
    W_{rKq}^{(k)}\,\theta_{Kq}
    }{
    \displaystyle\sum_{(K,q)\in\Omega_r^{\mathrm{GB}}}
    W_{rKq}^{(k)}
    }.
    \label{eq:results_region_rotation}
\end{equation}
Here $W_{rKq}^{(k)}\geq0$ is a smooth quadrature weight on frozen branch $k$.
It gives larger importance to points close to the selected grain boundary and
the center of the material patch, and smoothly emphasizes the larger rotation
angles within that patch. Consequently, $\omega_r$ characterizes the local
rotation concentration near the grain boundary without relying on the rotation
at a single mesh-dependent quadrature point.

The rotation mismatch is then defined by the normalized mean-squared difference
between the simulated and target regional rotations,
\begin{equation}
    J_{\omega}^{(k)}(\boldsymbol{\alpha})
    =
    \frac{1}{N_r}
    \sum_{r=1}^{N_r}
    \left(
    \frac{
    \omega_r(\boldsymbol{\alpha};\mathcal{B}_k)
    -\widehat{\omega}_r
    }{s_{\omega}}
    \right)^2,
    \label{eq:results_rotation_loss}
\end{equation}
where $\widehat{\omega}_r$ is the target rotation of region $r$ and
$s_{\omega}$ is the root-mean-square magnitude of the target rotations, with a
small positive lower bound used to avoid division by zero. For this example,
the complete objective in Eq.~\eqref{eq:method_frozen_objective} uses
$c_J=10^6$ and $\lambda_{\omega}=1$. 

Within each frozen remeshing branch, the objective is minimized using limited-memory Broyden–Fletcher–Goldfarb–Shanno (L-BFGS-B) method \cite{nocedal2006numerical}
with gradients supplied by JAX-based automatic differentiation. The
component-wise trust radius is $0.4$. A branch is refreshed when the normalized
distance $d_k$ from its anchor reaches $0.9$ of the valid radius or when its age
reaches four accepted inner iterations. 

\subsubsection{Results details}

\begin{figure}[htbp]
    \centering
    \includegraphics[scale=0.43]{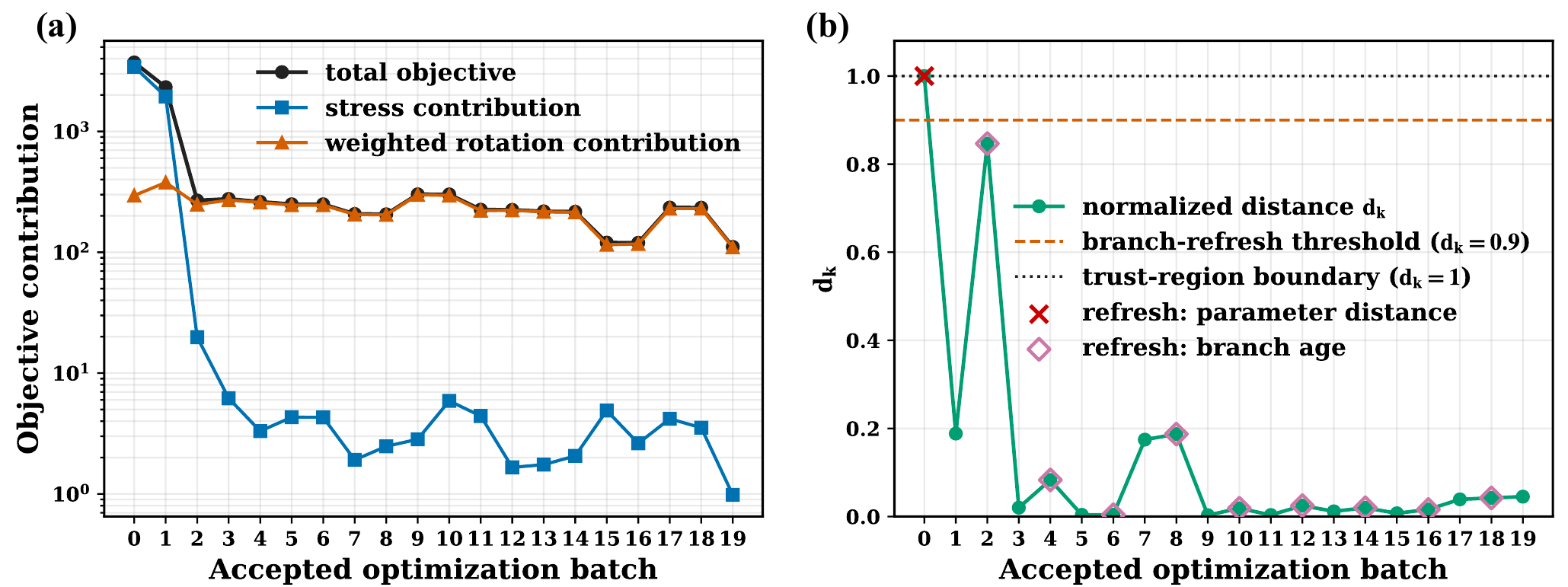}
    \caption{Optimization and frozen-branch management histories for the
    inverse-calibration problem. (a) Scaled total objective and its stress and
    weighted grain-boundary-rotation contributions versus accepted optimization
    batch. (b) Normalized distance $d_k$ from the accepted parameter to the
    current branch anchor. The dashed and dotted lines denote the branch-refresh
    threshold and trust-region boundary, respectively; the symbols identify
    refreshes caused by parameter distance and branch age.}
    \label{fig:calibration_optimization_history}
\end{figure}

Figure~\ref{fig:calibration_optimization_history} summarizes the convergence of
the frozen-branch calibration. As shown in
Fig.~\ref{fig:calibration_optimization_history}(a), the scaled objective is
reduced from $3.72\times10^{3}$ after the first accepted batch to
$1.11\times10^{2}$ after the twentieth batch, corresponding to a reduction of
approximately $97\%$. The decrease is particularly rapid during the first
three batches. The stress contribution decreases from $3.43\times10^{3}$ to
$9.83\times10^{-1}$ and therefore becomes negligible compared with the
rotation contribution in the later stage. This behavior indicates that the
macroscopic stress--strain response is identified first, whereas the remaining
optimization effort is primarily devoted to reducing the more sensitive local
rotation mismatch. The non-monotone changes in the total objective at several
batches are associated with branch regeneration: after the mesh sequence is
updated, the objective is evaluated on a new frozen remeshing branch and is
therefore not required to decrease monotonically across branch boundaries.

The corresponding parameter-trust history is shown in
Fig.~\ref{fig:calibration_optimization_history}(b). The first accepted update
reaches $d_k=1$ and triggers a refresh through the parameter-distance
criterion. Subsequent accepted parameters remain within the branch-valid box,
with $d_k<0.9$ in all later batches. Most later branch updates are instead
caused by the branch-age criterion. In total, 20 accepted optimization batches
are completed on 11 frozen branches. The fact that the accepted iterates remain
inside the trust-region boundary while branches are periodically regenerated
shows that the optimization can reuse a fixed mesh sequence locally without
allowing it to become indefinitely inconsistent with the evolving material
parameters.

\begin{figure}[htbp]
    \centering
    \includegraphics[scale=0.53]{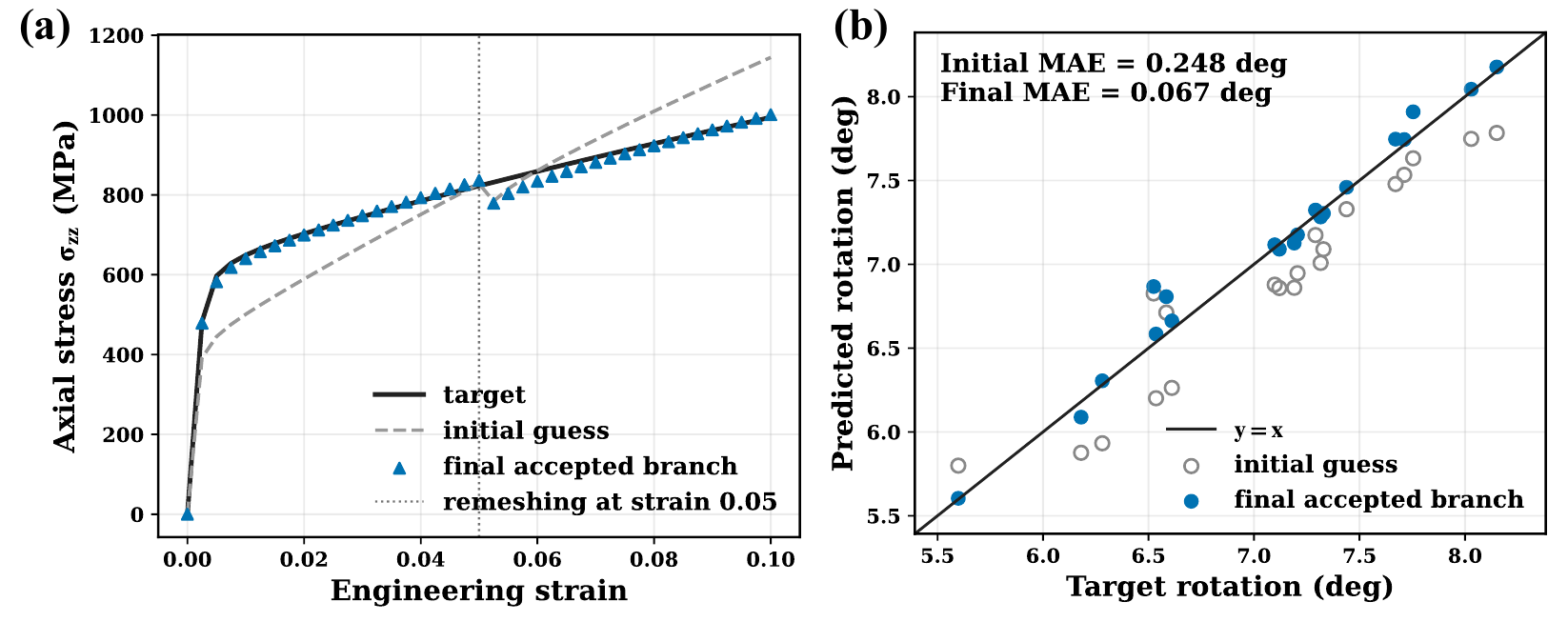}
    \caption{Comparison of the target responses with the initial and calibrated
    predictions. (a) Macroscopic axial stress--strain response; the vertical
    dotted line marks the remeshing point at an engineering strain of $0.05$.
    The final accepted response is shown by triangular markers. (b) Predicted
    versus target lattice rotations in the 20 selected grain-boundary regions.
    The solid line denotes exact agreement, $y=x$.}
    \label{fig:calibration_response_fit}
\end{figure}

The recovered target responses are compared in
Fig.~\ref{fig:calibration_response_fit}. The initial stress--strain curve shows
a substantial overprediction over most of the loading path. After calibration,
the stress values from the final accepted branch nearly coincide with the
target curve, including the response immediately before and after remeshing at
an engineering strain of $0.05$. Quantitatively, the normalized stress loss in
Eq.~\eqref{eq:results_stress_loss} decreases from $1.39\times10^{-2}$ for the
initial guess to $9.83\times10^{-7}$ for the final accepted branch. Thus, the
small discontinuity associated with remeshing does not prevent the optimizer
from recovering the complete macroscopic loading response.

Figure~\ref{fig:calibration_response_fit}(b) examines the 20 local
rotation targets independently of the stress response. The initial predictions
exhibit a systematic deviation from the identity line, while the calibrated
predictions concentrate much more closely around it. The mean absolute error is
reduced from $0.248^{\circ}$ to $0.067^{\circ}$, a reduction of approximately
$73\%$. This improvement confirms that the local grain-boundary observations
supply information that is not contained in the volume-averaged stress curve
alone. Some residual region-to-region errors remain because a single spatially
uniform parameter vector must simultaneously match all selected regions, whose
responses are also affected by heterogeneous grain orientations and local
neighborhoods.

\begin{figure}[htbp]
    \centering
    \includegraphics[scale=0.53]{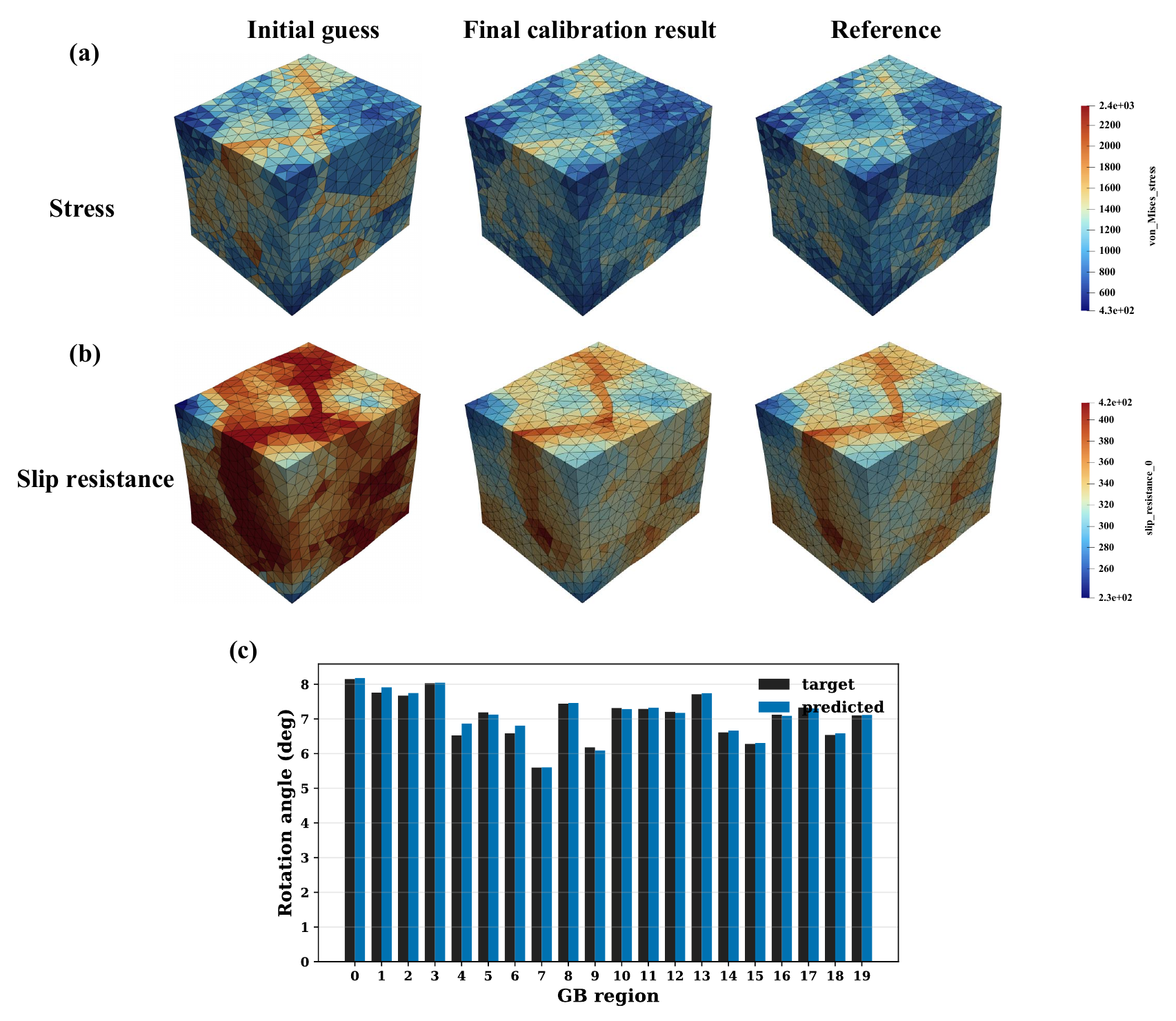}
    \caption{Final-state comparison of the initial guess, final calibration
    result, and reference solution. (a) Von Mises stress distributions.
    (b) Slip resistance distributions on slip system 0. Each row uses a common
    color range across the three columns. (c) Target and calibrated lattice
    rotations in the 20 selected grain-boundary regions.}
    \label{fig:calibration_field_comparison}
\end{figure}

The final-state field comparison in
Fig.~\ref{fig:calibration_field_comparison} provides a spatial assessment that
complements the scalar loss histories. The initial guess produces visibly
stronger and more extensive high-value regions in both the von Mises stress and
the slip resistance on slip system 0. After calibration, the magnitudes and
spatial organization of these fields move toward the reference solution: the
major stress concentrations and hardened bands are recovered at comparable
locations, and the excessive slip resistance in the initial result is
substantially reduced. The calibrated and reference fields are not expected to
agree pointwise. They are evaluated on independently generated remeshed meshes,
and the objective constrains the macroscopic stress history together with only
20 regional rotation measures rather than the complete stress and internal-
variable fields. Their agreement in the dominant spatial patterns is therefore
an additional out-of-objective validation rather than a directly enforced
result.

The regional comparison in Fig.~\ref{fig:calibration_field_comparison}(c)
further shows that the final predicted rotations closely follow the target
values across all selected grain-boundary regions, consistently with the
scatter plot in Fig.~\ref{fig:calibration_response_fit}(b). A final validation
calculation was additionally performed after regenerating the remeshing branch
at the calibrated parameter vector. It gives a normalized stress loss of
$1.93\times10^{-6}$ and a rotation loss of $1.83\times10^{-4}$, demonstrating
that the response improvement persists on a newly generated, parameter-
consistent branch and is not merely an artifact of the last frozen mesh
sequence.

Although the target data were generated synthetically from a known parameter
vector, the final accepted result from optimization batch 19 is
\begin{equation}
    \boldsymbol{\alpha}^{\star}
    =
    \left(
    2.5494,\,
    2.3118,\,
    1.3899,\,
    1.3059,\,
    2.5688,\,
    1.3899
    \right)^{T}.
    \label{eq:results_calibrated_parameters}
\end{equation}
The same parameter vector is used in the final validation calculation on the
newly generated branch. The calibrated coefficients do not recover every
reference coefficient in Table~\ref{tab:calibration_parameters} individually.
Instead, the optimization identifies a response-equivalent parameter set that
reproduces the prescribed global and local observables. This
result reflects correlations among the six constitutive parameters and the
limited identifiability provided by one loading path. Accordingly, the present
example validates the proposed differentiable remeshing calibration framework
for simultaneous recovery of macroscopic and selected local responses; unique
calibration of all constitutive parameters would require additional loading
paths or independent local measurements.

\section{Conclusions and discussions}
OmniRemesh presents a structure-driven remeshing framework for large deformation crystal plasticity that is built on the differentiable JAX-CPFEM solver. The method defines the target mesh on the current deformed configuration by combining grain-boundary proximity with internal-variable indicators derived from slip activity and hardening, then generates a new mesh, transfers state variables by grain-preserving nearest-neighbor mapping, and restores admissibility through an equilibrium projection step. In this way, the framework extends earlier remeshing ideas for polycrystal simulations toward locally adaptive, deformation-following refinement in CPFEM. The numerical examples show that the remeshed solutions remain close to fixed-mesh references in stress response and internal-field evolution, while providing substantially better mesh quality and sharper resolution of localized gradients at large strain. The large-deformation studies further show that the method preserves the overall deformation trend while improving the representation of grain-boundary and hotspot regions. For inverse analysis, the frozen-remeshing-branch strategy makes gradient-based calibration feasible despite topology changes, and the synthetic example shows that it can recover macroscopic stress histories together with selected local rotation observables with good accuracy. Overall, the results indicate that OmniRemesh is a practical route for combining adaptive remeshing, differentiable CPFEM, and inverse calibration in a single workflow

From an experimental perspective, the present synthetic calibration provides a controlled precursor to calibration against physical measurements. Hu et al. \citep{hu2026efficient} demonstrated such a workflow for wrought IN625 by constructing a polycrystalline RVE from EBSD-derived microstructural information and calibrating its constitutive parameters against an experimental tensile stress--strain curve. A similar experimental workflow could be adopted for OmniRemesh by replacing the synthetic targets with measured macroscopic loads and local deformation fields. The macroscopic objective could be defined using experimental stress--strain data, while local objectives could incorporate surface strain fields measured by digital image correlation or lattice rotations obtained from EBSD-based characterization. Comparisons among the original coarse-mesh simulation, the adaptively remeshed simulation, and the experimental measurements would then determine whether OmniRemesh improves not only numerical convergence at large strain, but also the prediction of experimentally observed strain localization and grain-scale heterogeneity. Independent loading paths or specimens should be reserved for validation rather than calibration to assess the transferability of the calibrated constitutive parameters.

At the same time, the present framework may be further improved in the following aspects. The remeshing operation itself remains non-differentiable, so the calibration relies on frozen mesh branches within a trust region and therefore provides only a piecewise-differentiable approximation. The nearest-neighbor state transfer is robust across grain boundaries, but it can still introduce local transfer errors, mesh dependence, and small stress discontinuities that must be corrected by equilibrium projection. The results also depend on user-chosen size-field parameters and indicator thresholds, and repeated remeshing, projection, and branch regeneration add nontrivial computational cost. In addition, the inverse example is synthetic and uses limited observables, so parameter identifiability remains incomplete under a single loading path. Future work will therefore develop more differentiable and less dissipative transfer operators, including hybrid interpolation/nearest-neighbor schemes, together with error estimators that adapt refinement criteria and trust-region updates more systematically. It will also extend the framework to broader and more demanding 3D applications, strengthen GPU/parallel efficiency, validate against experimental full-field measurements, and couple the remeshing workflow with phase-field, damage, and other microstructure-evolution models.

\section*{Acknowledgments} The research was partially supported by National Key R\&D Program of China  2022YFA1004500, and the Hong Kong Research Grants Council Collaborative Research Fund C7140-25G. 

\bibliographystyle{unsrtnat}
\bibliography{references}
\end{document}